# Is the LMA solar-neutrino solution ruled out by SN1987A data?[*]

David B. Cline

Department of Physics and Astronomy, Box 951647
University of California Los Angeles, Los Angeles, CA 90095-1647, USA

The development of new supernova neutrino detectors relies on the expected hard energy spectrum of the $\nu_\mu$ and $\nu_\tau$ emitted in the supernova. We show that SN1987A was sensitive to the large mixing angle (LMA) and "just so" solution to the solar neutrino problem. We review the previous analysis of the SN1987A data and propose a new analysis. The results of this analysis strongly disfavor the LMA solution, provided the $\nu_\mu$ and $\nu_\tau$ are hard as predicted

## 1. THE DEVELOPMENT OF THE NEW SUPERNOVA DETECTOR, OMNIS

Large water detectors are ideally matched for the detection of the process $\bar{\nu}_e + p \rightarrow e^+ + n$. The detection of the other flavors of neutrinos $\nu_\mu, \bar{\nu}_\mu, \nu_\tau, \bar{\nu}_\tau$ requires the observation of neutral current processes. Long ago [1,2], the author and G. Fuller and others proposed a method to carry out this detection by detecting the neutrons for

$$\nu_x + N \rightarrow \nu'_x + N'$$
$$\phantom{\nu_x + N \rightarrow \nu'_x +} \hookrightarrow n \ .$$

The concept has now been incorporated into the OMNIS detector proposal (Fig. 1) for the Carlsbad underground laboratory. The detection method relies on the expected harder $\nu_x$ spectrum as shown in Fig. 2.

## 2. NEUTRINO MIXING FROM SUPERNOVA NEUTRINOS

Because the $\nu_x$ spectrum is expected to be harder than the $\nu_e$ or $\bar{\nu}_e$ (Fig. 2), one signature for neutrino oscillation is to observe a hard component in the $\nu_e$ or $\bar{\nu}_e$ spectrum from the process $\nu_x \rightarrow \nu_e$ or $\bar{\nu}_x \rightarrow \bar{\nu}_e$ [3]. We can define the neutrino flux in the following way:

$$\langle \nu_e \rangle = (1-p)\langle \nu_e \rangle_o + p\langle \nu_x \rangle_o \ , \qquad (1)$$

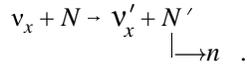

$$\langle \bar{\nu}_e \rangle = (1-\bar{p})\langle \bar{\nu}_e \rangle_o + \bar{p}\langle \bar{\nu}_x \rangle_o \ , \qquad (2)$$

where $\langle \nu_e \rangle_o$, $\langle \bar{\nu}_e \rangle_o$, $\langle \nu_x \rangle_o$, $\langle \bar{\nu}_x \rangle_o$ denote the unmixed neutrino spectra and $p, \bar{p}$ the mixing fraction.

In vacuum we can write $p = (1/2) \sin^2 2\theta$ and $\bar{p} = (1/2) \sin^2 2\theta$. Figure 3 shows the distorted $\bar{\nu}_e$ event spectrum for various values of $\bar{p}$. Thus, even a small mixing ($\bar{p} = 0.2$) causes an appreciable event spectrum distortion at high energy and should be readily detected in the next supernova event.

## 3. EARLY ANALYSES OF SN1987A DATA

As is well known, there were 20 events recorded in SN1987A: 12 by the Kamiokande detector [4] and 8 by the IMB detector [5]. We first turn to the initial analysis of L. Krauss [6].

First we comment on the Kamiokande and IMB event populations:

1. The IMB detector had a strong bias against low energy events.
2. The mass of the IMB detector was about three times larger than the Kamiokande detector and thus was more sensitive to higher energy neutrino events that are less probable.
3. The Kamiokande detector had excellent low energy properties, as was later demonstrated by the observation of solar neutrinos.
4. Some of the pmt's for the IMB detector were not operational during the recording of SN1987A events.

---

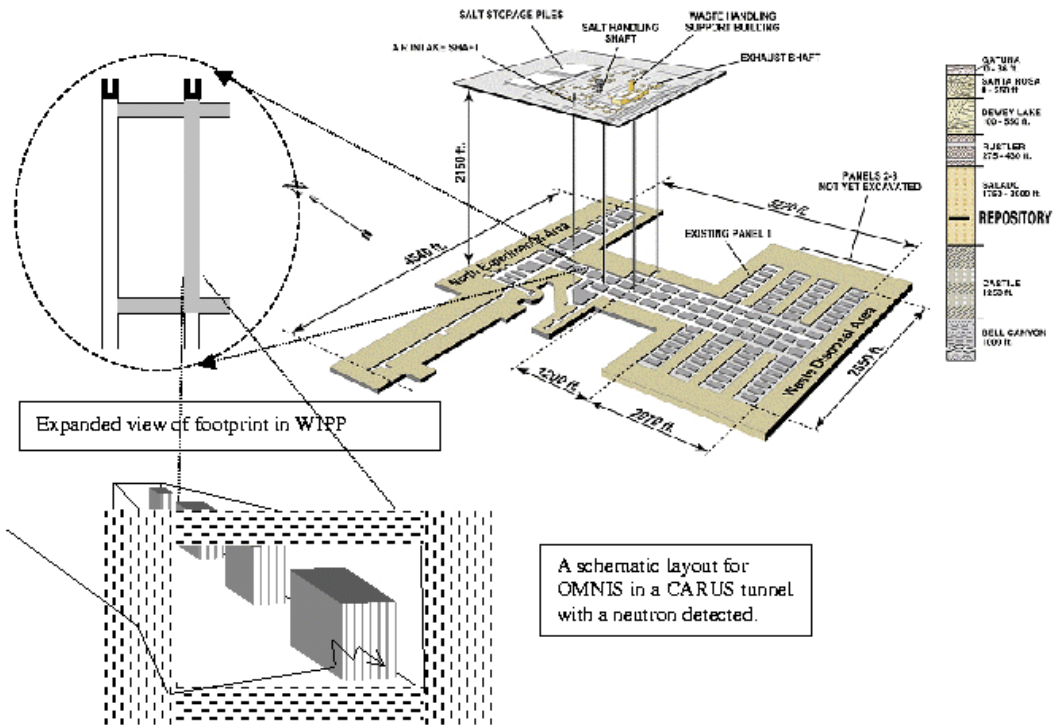

Figure 1. Schematic Layout for OMNIS in a CARUS tunnel with a neutron detected.

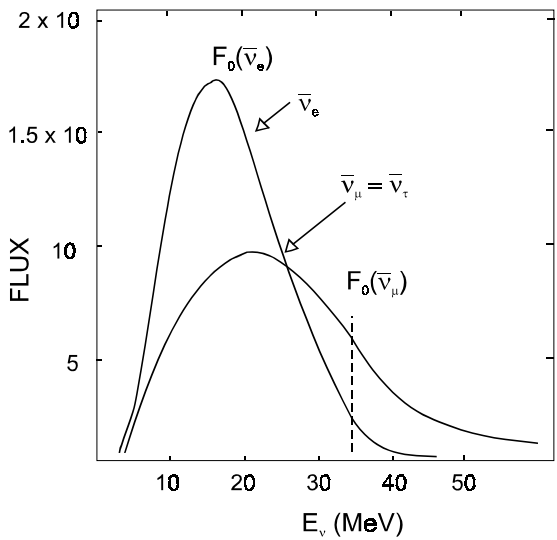

Figure 2. Expected $\bar{\nu}_e$ and $\bar{\nu}_x$ spectra from an SNII explosion.

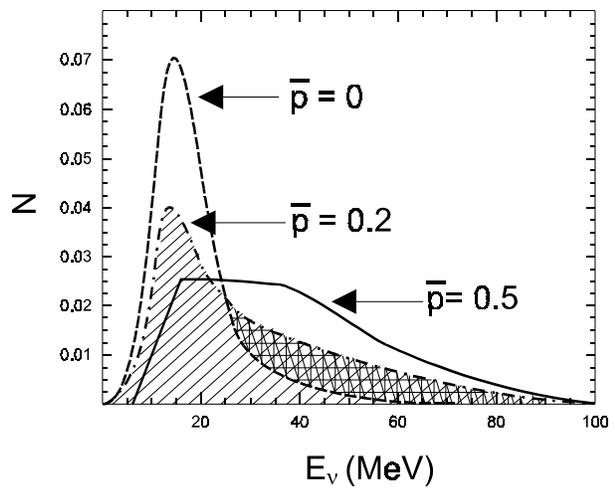

Figure 3. Events distribution expected for different values of the mixing $\bar{p}$.

In the Krauss analysis, an attempt was made to incorporate both populations of events by correcting the model as shown in Fig. 4. While this showed an acceptable fit, the mean temperature of the distribution was 4.5 MeV, which was higher than would be expected normally.

Smirnov, Spergel, and Bahcall tried another analysis [7]. They fit the combined Kamiokande and IMB data with a model that allowed for $v_x \rightarrow v_e$ mixing as given by formula (2). They did not correct for the obvious differences in the Kamiokande and IMB event population (threshold, different detector, masses, possible differences in detection efficiency, etc.) but just added up the integral of the total event energies. They gave an exhaustive discussion of the different types of temperature distributions that may occur for the different neutrino flavors from the supernova emission. They expressed their results as a function of the mean energy of the $v_x$ neutrinos. Since most models gave this value to be 22 MeV or greater, we use that value in the results shown here. In Fig. 5, we have replotted the results of this analysis for the 95% confidence level reported in the paper. Note that most of the LMA and all of the vacuum oscillation (or "just so") is excluded [8].

One can be critical of this analysis due to the fact that no attempt was made to correct for the different experimental conditions in the Kamiokande and IMB experiments. However, these results may well be a conservative lower limit, since corrections for the experimental differences will decrease the impact of the high energy events in IMB, as shown by the Krauss analysis.

## 4. A NEW ANALYSIS OF THE SN1987A DATA FOR NEUTRINO MIXING

Because of the problems of comparing the two populations of events illustrated in points 1 through 4 above, we propose that a sensible analysis should use the data set with the least bias. Based on the Kamiokande data alone, while this set has no event with an energy above 40 MeV, there is no reason why the detector would not have recorded such events had they been produced. In Fig. 6, we show the Kamiokande data and the predictions of neutrino mixing (Fig. 3). The case $\bar{p} = 1/2$ is excluded to at least 99% confidence level. Even with a lower statistical sample, the conclusions of this analysis are as powerful as those of Smirnov, Spergel, and Bahcall. Table 1 gives the Kolmogurov test for these data [9]. We believe these results largely exclude the LMA solar neutrino solution.

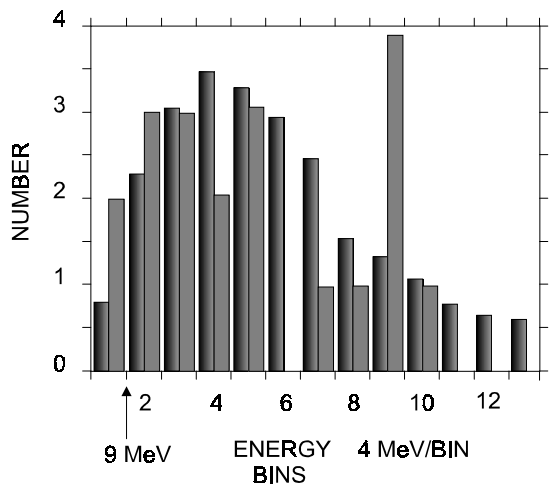

Figure 4.  Krauss analysis of SN1987A data. The hatched events are the combined data and the solid blocks are the results of a model that incororates the effects of the detectors, etc. [6]. Neutrino mixing was not assumed.

## 5. SUMMARY

Our conclusions are that either (1) the SNII model with a high-energy $v_x$ spectrum is not correct, or (2) very likely the LMA solar neutrino solution is excluded. We note that the IMB data indicates that a supernova II produces high-energy $\bar{v}_e$ events, so it is very likely that the standard predictions of the supernova II models are correct.

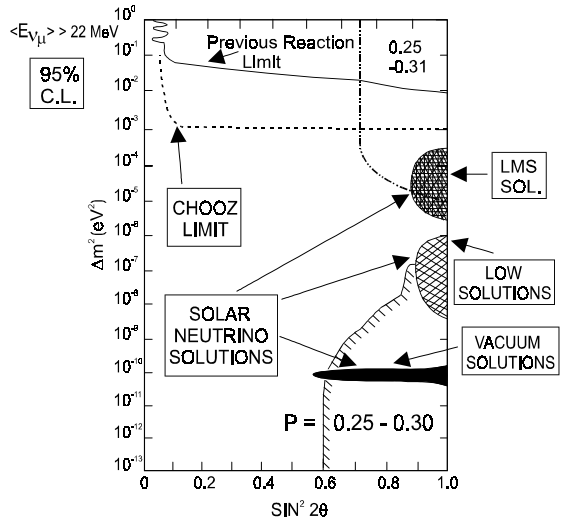 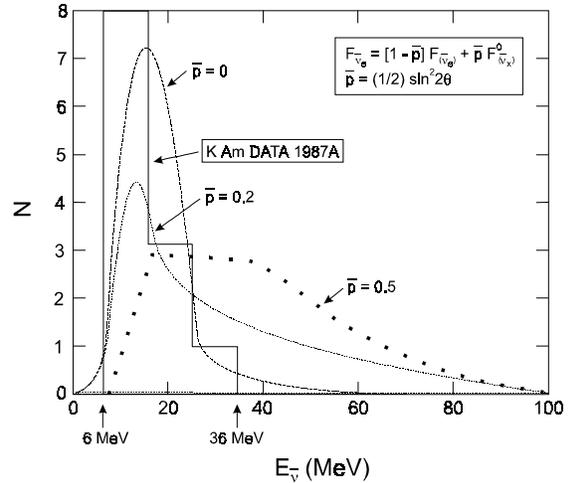

Figure 5. Limits on $\bar{\nu}_e \rightarrow \bar{\nu}_\mu$ from SN1987A (from Ref. [7]).

Figure 6. Comparison of the Kamiokande data with the neutrino oscillation models.

Table 1. Kolmogurov Test for the Model and data in Fig. 6.

| Parameter, $\bar{p}$ | Probability of Hypothesis (%) | Confidence Level (%) | |
|---|---|---|---|
| 0 | 58 | 42 | |
| 0.2 | 3.6 | 96.4 | excluded |
| 0.5 | 0.02 | > 99 | excluded |

## 6. ACKNOWLEDGEMENT

I wish to thank G. Fuller and Yu Smirnov for comments on this analysis.